\documentclass[preprint,preprintnumbers,amsmath,amssymb]{revtex4-1}
\usepackage{amsfonts}    
\usepackage{amssymb}
\usepackage{latexsym}
\usepackage{eepic}
\usepackage{graphicx}
\usepackage[usenames,dvipsnames]{color}

\usepackage[active]{srcltx}

\newcommand{\pa}{\partial}

\newcommand{\si}{\sigma}

\newcommand{\ta}{\tau}

\newcommand{\Om}{\Omega}
\newcommand{\om}{\omega}
\newcommand{\de}{\delta}

\newcommand{\De}{\Delta}

\newcommand{\rar}{\rightarrow}
\newcommand{\lrar}{\leftrightarrow}

\newcommand{\non}{\nonumber}

\begin{document}

\title{Three-body problem in 3D space: ground state, (quasi)-exact-solvability}

\author{Alexander V Turbiner\\[8pt]
Instituto de Ciencias Nucleares, UNAM, M\'exico DF 04510, Mexico\\
and\\
IHES, Bures-sur-Yvette, France,\\[8pt]
turbiner@nucleares.unam.mx\\[8pt]
     Willard Miller, Jr.\\[8pt]
School of Mathematics, University of Minnesota, \\
Minneapolis, Minnesota, U.S.A.\\[8pt]
miller@ima.umn.edu\\
[10pt]
and \\[10pt]
Adrian~M~Escobar-Ruiz,\\[8pt]
Instituto de Ciencias Nucleares, UNAM, M\'exico DF 04510, Mexico\\
and\\
School of Mathematics, University of Minnesota, \\
Minneapolis, Minnesota, U.S.A.\\[8pt]
mauricio.escobar@nucleares.unam.mx}

\begin{abstract}
We study aspects of the quantum and classical  dynamics of a $3$-body system in 3D space with interaction depending only on mutual distances.
The study is restricted to solutions in the space of relative motion which are functions of mutual distances only. It is shown that the ground state (and some other states) in the quantum case and the planar trajectories in the classical case are of this type. The quantum (and classical) system for which
these states are eigenstates is found and its Hamiltonian is constructed. It corresponds to a three-dimensional quantum particle moving in a curved space with special metric.
The kinetic energy of the system has a hidden $sl(4,R)$ Lie (Poisson) algebra structure, alternatively, the hidden algebra $h^{(3)}$ typical for the $H_3$ Calogero model. We find an exactly solvable three-body generalized harmonic oscillator-type potential as well as a quasi-exactly-solvable three-body
sextic polynomial type potential;  both models have an extra integral.

\end{abstract}

\maketitle

\newpage

\section*{Introduction}

The Hamiltonian for $3$-body quantum system of $3$-dimensional particles with translation-invariant potential, which depends on relative distances between particles only, is of the form,
\begin{equation}
\label{Hgen}
   {\cal H}\ =\ -\sum_{i=1}^3 \frac{1}{2m_i}\De_i^{(3)}\ +\  V(r_{12},\,r_{13},\,r_{23})\ ,\
\end{equation}
with coordinate vector of $i$th particle ${\bf r}_i \equiv {\bf r}^{(3)}_i=(x_{i,1}\,, x_{i,2}\,,x_{i,3})$\,, where
\begin{equation}
\label{rel-coord}
r_{ij}=|{\bf r}_i - {\bf r}_j|\ ,
\end{equation}
is the (relative) distance between particles $i$ and $j$. We consider the case when all masses are assumed to be equal: $m_i=m=1$. In this case the kinetic energy operator is $-\frac{\De^{(9)}}{2}$\, where $\De^{(9)}$ is nine-dimensional Laplacian.
The number of relative distances is equal to the number of edges of the triangle
formed by taking the body positions as vertices. We call this triangle the {\it triangle of interaction}. Here, $\De_i^{(3)}$ is the $3$-dimensional Laplacian,
\[
     \De_i^{(d)}\ =\ \frac{\pa^2}{\pa{{\bf r}_i} \pa{{\bf r}_i}}\ ,
\]
associated with the $i$th body.
 The configuration space for ${\cal H}$ is ${\bf R}^{9}$.
The center-of-mass motion described by vectorial coordinate
\[
    {\bf R}_{_0} \ =\ \frac{1}{{\sqrt 3}}\,\sum_{k=1}^{3} {\bf r}_{_k}\ ,
\]
can be separated out; this motion is described by a $3$-dimensional plane wave.

The spectral problem is formulated in the space of relative motion ${\bf R}_r \equiv {\bf R}^{6}$; it is of the form,
\begin{equation}
\label{Hrel}
   {\cal H}_r\,\Psi(x)\ \equiv \ \bigg(- \frac{1}{2}\De_r^{(6)} + V(r_{12},\,r_{13},\,r_{23})\bigg)\, \Psi(x)\ =\ E \Psi(x)\ ,\
   \Psi \in L_2 ({\bf R}_r)\ ,
\end{equation}
where $\De_r^{(6)}$ is the flat-space Laplacian in the space of relative motion. If the space of relative motion ${\bf R}_r$
is parameterized by two, $3$-dimensional vectorial Jacobi coordinates
\[
     {\bf r}^{(F)}_{j} \ = \ \frac{1}{\sqrt{j(j+1)}}\sum_{k=1}^j k\,({\bf r}_{k+1} - {\bf r}_{{k}})\ ,
        \qquad\qquad j=1,2\ ,
\]
the flat-space $6$-dimensional Laplacian in the space of relative motion becomes diagonal
\begin{equation}
\label{Dflat}
       \De_r^{(6)}\ =\ \frac{\pa^2}{\pa{{\bf r}_i^{(F)}} \pa{{\bf r}_i^{(F)}}}\ .
\end{equation}

\newpage

{\large \it Observation:}
\begin{quote}
  There exists a family of the eigenstates of the Hamiltonian (\ref{Hgen}), including the ground state,
  which depends on three relative distances $\{r_{ij}\}$ only\, .
\end{quote}

\bigskip
\noindent
Our primary goal is to find the differential operator in the space of relative distances $\{r_{ij}\}$ for which these states are eigenstates. In other words, to find a differential equation depending only on $\{r_{ij}\}$ for which these states are solutions. This  implies a study of the evolution of the triangle of interaction.


\section{Generalities}

As a first step let us change variables in the space of relative motion ${\bf R}_r$ : $({\bf r}^{(F)}_{j}) \lrar (r_{ij}, \Om)$,
where the number of (independent) relative distances $r_{ij}$ is equal to 3
and $\Om$ is a collection of three angular variables. Thus, we split ${\bf R}_r$ into a sum of the space of relative distances
${\bf \tilde R}$ and a space parameterized by angular variables, essentially those on the  sphere $S^3$. There are known several ways to
introduce variables in ${\bf R}_r$: the perimetric  coordinates by Hylleraas \cite{Hylleraas}, the scalar products of vectorial Jacobi
coordinates ${\bf r}^{(F)}_{j}$ \cite{Gu} and the relative (mutual) distances $r_{ij}$ (see e.g. \cite{Loos}). We follow the last one.
In turn, the angular variables are introduced as the two Euler angles on the $S^2$ sphere defining the normal to the interaction plane (triangle) and the azimuthal angle of rotation of the interaction triangle around its barycenter, see e.g. \cite{Gu}.

A key observation is that in new coordinates $(r_{ij}, \Om)$ the flat-space Laplace operator (the kinetic energy operator)
in the space of relative motion ${\bf R}_r$ takes the form of the sum of two the second-order differential operators
\begin{equation}
\label{addition}
    \frac{1}{2}\De_r^{(6)}\ =\ {\De_R}(r_{ij}) + {\tilde \De} (r_{ij}, \Om, \pa_{\Om})\ ,
\end{equation}
where the first operator depends on relative distances {\it only},  while the second operator depends on angular derivatives in such a way
that it annihilates any angle-independent function,
\[
  {\tilde \De} (r_{ij}, \Om, \pa_{\Om})\, \Psi(r_{ij})\ =\ 0\ .
\]

If we look for angle-independent solutions of (\ref{Hrel}),
the decomposition (\ref{addition}) reduces the general spectral problem (\ref{Hrel})
to a particular spectral problem
\begin{equation}
\label{Hrel-Mod}
   {\tilde {\cal H}}_R\,\Psi(r_{ij}) \equiv \ \bigg(- {\De_R}(r_{ij}) + V(r_{12},\,r_{13},\,r_{23})\bigg)\, \Psi(r_{ij})=E\Psi(r_{ij})\ ,\
   \Psi \in L_2 ({\bf \tilde R})\ ,
\end{equation}
where ${\bf \tilde R}$ is the space of relative distances. Surprisingly,
one can find the gauge factor $\Gamma(r_{ij})$ such that the operator ${\De_R}(r_{ij})$ takes the form of the Schr\"odinger operator,
\begin{equation}
\label{DLB}
     \Gamma^{-1}\,{\De_R}\,(r_{ij})\, \Gamma\ =\ {\De_{LB}}(r_{ij}) - {\tilde V}(r_{ij})\ \equiv
      -{\tilde H}_R \ ,
\end{equation}
where $\De_{LB}$ is the Laplace-Beltrami operator with contravariant metric $g^{ij}$, in general, on some non-flat, (non-constant curvature) manifold.
It makes sense of the kinetic energy.
Here ${\tilde V}(r_{ij})$ is the effective potential. The potential ${\tilde V}$ becomes singular at the boundary of the configuration space, where the determinant $D=\det g^{ij}$ vanishes. The operator ${\tilde H}_R$ is Hermitian with measure $D^{-\frac{1}{2}}$. Eventually, we arrive at the spectral problem for the Hamiltonian
\begin{equation}
\label{Hrel-final}
   {H}_R\ =\ -{\De_{LB}}(r_{ij}) + V(r_{ij}) + {\tilde V}(r_{ij})\ .
\end{equation}
Following the {\it de-quantization} procedure of replacement of the quantum momentum (derivative) by the classical momentum
\[
      -i\,\pa\ \rar\ p\ ,
\]
one can get a classical analogue of (\ref{Hrel-final}),
\begin{equation}
\label{Hrel-Cl-final}
   {H}^{(c)}_R\ =\ g^{\mu \nu} p_{\mu} p_{\nu} + V(r_{ij}) + {\tilde V}(r_{ij})\ .
\end{equation}
It describes the motion of 3-dimensional rigid body with tensor of inertia $(g^{\mu \nu})^{-1}$\,.

The Hamiltonians (\ref{Hrel-final}), (\ref{Hrel-Cl-final}) are the main objects of study of this paper.

\section{Three-body case: concrete results}

After straightforward calculations the operator ${\De_R}(r_{ij})$ in decomposition (\ref{addition}) is found to be
\begin{equation}
\label{addition3-3r}
   2\, {\De_R}(r_{ij})\ =\ \bigg[\ 2\,(\pa^{2}_{r_{12}} +\pa^{2}_{r_{23}}+\pa^{2}_{r_{13}})
+ \frac{4}{r_{12}}\,\pa_{r_{12}}  +  \frac{4}{r_{23}}\,\pa_{r_{23}} + \frac{4}{r_{13}}\,\pa_{r_{13}}
\end{equation}
\[
  + \frac{r_{12}^2-r_{13}^2+r_{23}^2}{r_{12} r_{23}}\,\pa_{r_{12}}\pa_{r_{23}}  + \frac{r_{12}^2+r_{13}^2-r_{23}^2}{r_{12} r_{13}}\,\pa_{r_{12}}\pa_{r_{13}} +  \frac{r_{13}^2+r_{23}^2-r_{12}^2}{r_{13} r_{23}}\,\pa_{r_{23}}\pa_{r_{13}} \ \bigg]\ ,
\]
cf. e.g. \cite{Loos}.
It does not depend on the choice of the angular variables $\Om$. Its configuration space is
\begin{equation}
\label{CFr}
 0 < r_{12},r_{13},r_{23} < \infty,\quad  r_{23} < \ r_{12} + r_{13},\quad r_{13}< r_{12}+r_{23},\quad r_{12}< r_{13}+r_{23}.
\end{equation}
In the space with Cartesian coordinates $(x,y,z)=(r_{12},r_{13},r_{23})$ the configuration space lies in the first octant and is the interior
of the inverted  tetrahedral-shaped object with base at infinity, vertex at the origin and edges $(t,t,2t)$, $(t,2t,t)$ and $(2t,t,t)$, $0\leq t<\infty$.

Formally, the operator (\ref{addition3-3r}) is invariant under reflections $Z_2 \oplus Z_2 \oplus Z_2$,
\[
r_{12} \rightarrow -r_{12}  \ ,\qquad  r_{13} \Leftrightarrow -r_{13} \ ,\qquad  r_{23} \Leftrightarrow -r_{23} \ ,
\]
and w.r.t. $S_3$-group action. If we introduce new variables,
\begin{equation}
\label{rho}
r_{12}^2\ =\ \rho_{12}\ ,\ r_{13}^2\ =\ \rho_{13}\ ,\ r_{23}^2\ =\ \rho_{23}\ ,
\end{equation}
the operator (\ref{addition3-3r}) becomes algebraic,
\[
  {\De_R}(\rho_{ij})\ =\ 4(\rho_{12} \pa^2_{\rho_{12}} + \rho_{13} \pa^2_{\rho_{13}} +\rho_{23} \pa^2_{\rho_{23}}) + 6 (\pa_{\rho_{12}} + \pa_{\rho_{13}}+ \pa_{\rho_{23}}) +
\]
\begin{equation}
  2 \bigg((\rho_{12} + \rho_{13} - \rho_{23})\pa_{\rho_{12}}\pa_{\rho_{13}}\ +
          (\rho_{12} + \rho_{23} - \rho_{13})\pa_{\rho_{12}}\pa_{\rho_{23}}\ +
          (\rho_{13} + \rho_{23} - \rho_{12})\pa_{\rho_{13}}\pa_{\rho_{23}}
    \bigg)  \ .
\label{addition3-3rho}
\end{equation}
From (\ref{CFr}) and (\ref{rho}) it follows that the corresponding configuration space in $\rho$ variables is given by the conditions
\[
0 < \rho_{12},\rho_{13},\rho_{23} < \infty,\
{\rho}_{23} <  (\sqrt{{\rho}_{12}} + \sqrt{{\rho}_{13}})^2,\ {\rho}_{13} < (\sqrt{{\rho}_{12}} + \sqrt{{\rho}_{23}})^2,\ {\rho}_{12} < \ (\sqrt{{\rho}_{13}} + \sqrt{{\rho}_{23}})^2.
\]
We remark that
\begin{equation}
\label{CFrho}
\quad
\rho _{12}^2+\rho _{13}^2+\rho _{23}^2 -2 \rho _{12} \rho _{13}- 2 \rho _{12} \rho _{23}-
       2 \rho _{13} \rho _{23}\ <\ 0   \ ,
\end{equation}
because the left-hand side (l.h.s.) is equal to
$$-(r_{12}+r_{13}-r_{23})(r_{12}+r_{23}-r_{13})(r_{13}+r_{23}-r_{12})(r_{12}+r_{13}+r_{23})$$
and conditions (\ref{CFr}) should hold. Therefore, following the Heron formula, l.h.s. is proportional to the square of the area of the triangle of interaction $S^2_{\triangle}$\,.

The associated contravariant metric for the operator ${\De_R}(\rho_{ij})$ defined by coefficients in front of second derivatives is remarkably simple
\begin{equation}
\label{gmn33-rho}
 g^{\mu \nu}(\rho)\ = \left|
 \begin{array}{ccc}
 4\rho_{12} & \rho_{12} + \rho_{13} - \rho_{23} & \rho_{12} + \rho_{23} - \rho_{13} \\
            &                                   &                                   \\
 \rho_{12} + \rho_{13} - \rho_{23} & 4\rho_{13} & \rho_{13} + \rho_{23} - \rho_{12} \\
            &                                   &                                   \\
 \rho_{12} + \rho_{23} - \rho_{13} & \rho_{13} + \rho_{23} - \rho_{12} & 4\rho_{23}
 \end{array}
               \right| \ ,
\end{equation}
it is linear in $\rho$-coordinates(!) with factorized determinant
\begin{equation}
\label{gmn33-rho-det}
\det g^{\mu \nu}\ =\ - 6\left(\rho _{12}+\rho _{13}+\rho _{23}\right)
                     \left(\rho _{12}^2+\rho _{13}^2+\rho _{23}^2 -2 \rho _{12} \rho _{13}-
                     2 \rho _{12} \rho _{23}-2 \rho _{13} \rho _{23}\right) \equiv D> 0\ ,
\end{equation}
and is positive definite. It is worth noting a remarkable factorization property of the determinant
\[
D\ =\ 6\,(r_{12}^2+r_{13}^2+r_{23}^2)\ \times
\]
\[
(r_{12}+r_{13}-r_{23})(r_{12}+r_{23}-r_{13})(r_{13}+r_{23}-r_{12})(r_{12}+r_{13}+r_{23})\ =
\]
\[
   =\ 96\, P \ S^2_{\triangle}\ ,
\]
where $P=r_{12}^2+r_{13}^2+r_{23}^2$ - the sum of squared of sides of the interaction triangle.

The determinant can rewritten in terms of elementary symmetric polynomials $\si_{1,2}$,
\begin{equation}
\label{gammas}
\begin{aligned}
&  \ta_1=\si_1(\rho _{12},\,\rho _{13},\,\rho _{23}) \ = \ \rho _{12}+\rho _{13}+\rho _{23}\ ,
\\ & \ta_2=\si_2(\rho _{12},\,\rho _{13},\,\rho _{23}) \ = \  \rho _{12} \,\rho _{13}+
 \rho _{12} \,\rho _{23}+ \rho _{13}\, \rho _{23} \ ,
\\
&  \ta_3=\si_3(\rho _{12},\,\rho _{13},\,\rho _{23}) \ = \ \rho _{12}\rho _{13}\rho _{23}\ ,
\end{aligned}
\end{equation}
which are invariant w.r.t. $S_3$-group action, as follows,
\begin{equation}
\label{gmn33-rho-det-gamma}
      D\ =\ 6\, \ta_1\ (4\ta_2-\ta_1^2)\ ,
\end{equation}
where $$16 S^2_{\triangle}=(4\ta_2-\ta_1^2)\ ,$$ in terms of the elementary symmetric polynomials $\ta_{1,2}$.
When $\det g^{\mu \nu}=0$, hence, either $\ta_1=0$, or $\ta_1^2 = 4 \ta_2$ - it defines the boundary of the configuration space, see (\ref{CFrho}).

It can be shown that there exists the 1st order symmetry operator
\begin{equation}
L_1 \ = \  (\rho_{13}-\rho_{23})\pa_{\rho_{12}} + (\rho_{23}-\rho_{12})\pa_{\rho_{13}} + (\rho_{12}-\rho_{13})\pa_{\rho_{23}} \ ,
\label{integral}
\end{equation}
for the operator (\ref{addition3-3rho}),  $$[{\De_R}(\rho_{ij})\ ,\ L_1]=0\ .$$
Here, $L_1$ is an algebraic operator, which is anti-invariant under the $S_3$-group action.
The existence of the symmetry operator $L_1$ implies that in the space of relative distances one variable
can be separated out in (\ref{addition3-3rho}).

Set
\begin{equation}
\label{wcoords1}
     w_1\ =\ \rho_{12}+\rho_{13}+\rho_{23}\quad ,\quad
     w_2\ =\ 2\,
       \sqrt{\rho_{12}^2+\rho_{13}^2+\rho_{23}^2-\rho_{12}\rho_{13}-\rho_{12}\rho_{23} -\rho_{13}\rho_{23}}
       \quad ,
\end{equation}
 where $w_2=2\sqrt{(\ta_1^2 - 3 \ta_2)}$ as well\,, which are
invariant under the action of $L_1$, and
\[
   w_3=\frac{\sqrt{3}}{9}\left({\rm sgn}\,(\rho_{23}-\rho_{13})\arcsin(\frac{2\rho_{12}-\rho_{23}-\rho_{13}}{w_2})
   +{\rm sgn}(\rho_{13}-\rho_{12})\,\arcsin(\frac{2\rho_{23}-\rho_{13}-\rho_{12}}{w_2})\right.
\]

\begin{equation}
\label{wcoords2}\left.
   +{\rm sgn}(\rho_{12}-\rho_{23})\,\arcsin(\frac{2\rho_{13}-\rho_{23}-\rho_{12}}{w_2})-\frac{3\pi}{4}\right),
\end{equation}
with ${\rm sgn}(x)=\frac{x}{|x|}$ for nonzero $x$. These coordinates are invariant under a cyclic permutation of the indices on the $\rho_{jk}$: $1\to 2\to 3\to 1$. Under a transposition of exactly two indices, see e.g. $(12), (3)$\,, we see that $w_1,w_2$ remain invariant,
and $w_3 \to -w_3-\frac{\sqrt{3}\pi}{6}$.
Expressions for $w_3$ vary, depending on which of the 6 non-overlapping regions of
$(\rho_{12}, \rho_{13}, \rho_{23})$ space we choose to evaluate them:
\begin{enumerate}
 \item \[ (a):\ \rho_{23}>\rho_{13}>\rho_{12}\ ,\quad (b):\ \rho_{13}>\rho_{12}>\rho_{23}\ ,\quad (c):\ \rho_{12}>\rho_{23}>\rho_{13}\ ,\]
\item   \[(d):\ \rho_{13}>\rho_{23}>\rho_{12}\ ,\quad (e):\ \rho_{12}>\rho_{13}>\rho_{23}\ ,\quad (f):\ \rho_{23}>\rho_{12}>\rho_{13}\ ,\]
\end{enumerate}

The regions in class 1 are related by cyclic permutations, as are the regions in class 2.
We  map between regions by a transposition. Thus it is enough to evaluate $w_3$ in the region $(a):\ \rho_{23}>\rho_{13}>\rho_{12}$. The other 5 expressions will then follow from the
permutation symmetries. In this case we have
\[ (a):\  w_3=-\frac{\sqrt{3}}{9}\arcsin\left[\frac{2\sqrt{2}}{w_2^3}((2-\sqrt{3})\rho_{13}-\rho_{23}+(\sqrt{3}-1)\rho_{12})\times\right.\]
\[\left.(2\rho_{23}
 -(1+\sqrt{3})\rho_{13}+(\sqrt{3}-1)\rho_{12})((2+\sqrt{3})\rho_{12}-(1+\sqrt{3})\rho_{13}-\rho_{23})\right].\]
 (The special cases where exactly two of the $\rho_{jk}$ are equal can be obtained from these results by continuity. Here, $w_3$ is a single-valued differentiable function of
 $\rho_{12},\rho_{13},\rho_{23}$ everywhere in the physical domain (configuration space), except for the points $\rho_{12}=\rho_{13}=\rho_{23}$ where it is undefined.)

In  these coordinates, the operators (\ref{integral}) and  (\ref{addition3-3rho}) take the form
\[
 L_1(w)  \ = \pa_{w_3}\ ,
\]
\[
 \De_R(w) \ =\ \ 6\,w_1\pa_{w_1}^2\ +\ 6\,w_1\pa_{w_2}^2+2\,\frac{w_1}{w_2^2}\pa_{w_3}^2
 \ +\ 12\,w_2\pa_{w_1w_2}^2\ +\ 18\,\pa_{w_1}
 \ +\ 6\,\frac{w_1}{w_2}\pa_{w_2} \ ,
\]
respectively. It is evident that for the $w_3$-independent potential
\[
    V(w_1,\,w_2; w_3) \ = \  g(w_1,\,w_2) \ ,
\]
the operator $L_1$ is still an integral for an arbitrary function $g$:
\ $[L_1(w) , -\De_R(w) + g(w_1,\,w_2)] = 0 $. This property of integrability allows us
the separation of the variable $w_3$ in the spectral problem
\[
   [-\De_R(w) + g(w_1,\,w_2)]\Psi \ =\ E \Psi\ ,
\]
where $\Psi = \psi(w_1,\,w_2)\,\xi(w_3)$ is defined by the differential equations,
\begin{equation}
\label{eq-xi}
   \pa_{w_3} \xi\ =\ i\,m \xi\ ,\quad \xi\,=\,e^{i m \theta}\ ,
\end{equation}
\begin{equation}
\label{eq-psi}
  \left( 6\,w_1\pa_{w_1}^2\ +\ 6\,w_1\pa_{w_2}^2
 \ +\ 12\,w_2\pa_{w_1w_2}^2\ +\ 18\,\pa_{w_1}\ +\ 6\,\frac{w_1}{w_2}\pa_{w_2}
  - 2\,m^2\,\frac{w_1}{w_2^2}  - g(w_1,\,w_2) \right) \psi\ =\ -E \psi\ .
\end{equation}

Both operators (\ref{addition3-3rho}) and (\ref{integral}) are $sl(4,{\bf R})$-Lie algebraic - they can be rewritten in terms of the
generators of the maximal affine subalgebra $b_4$ of the algebra $sl(4,{\bf R})$, see e.g. \cite{Turbiner:1988,Turbiner:2016}
\begin{eqnarray}
\label{sl4R}
 {\cal J}_i^- &=& \frac{\pa}{\pa u_i}\ ,\qquad \quad i=1,2,3\ , \non  \\
 {{\cal J}_{ij}}^0 &=&
               u_i \frac{\pa}{\pa u_j}\ , \qquad i,j=1,2,3 \ , \\
 {\cal J}^0(N) &=& \sum_{i=1}^{3} u_i \frac{\pa}{\pa u_i}-N\, , \non \\
 {\cal J}_i^+(N) &=& u_i {\cal J}^0(N)\ =\
    u_i\, \left( \sum_{j=1}^{3} u_j\frac{\pa}{\pa u_j}-N \right)\non ,
       \quad i=1,2,3\ ,
\end{eqnarray}
where $N$ is parameter and
\[
 u_1\equiv\rho_{12}\ ,\qquad u_2\equiv\rho_{13}\ , \qquad u_3\equiv\rho_{23} \ .
\]
If $N$ is non-negative integer, a finite-dimensional representation space occurs,
\begin{equation}
\label{P3}
     {\cal P}^{(3)}_{N}\ =\ \langle u_1^{p_1} u_2^{p_2} u_3^{p_3} \vert \ 0 \le p_1+p_2+p_3 \le N \rangle\ .
\end{equation}
Explicitly, these operators look as
\begin{equation}
\label{HRex}
\De_R({\cal J}) \ = \ 4(\, {\cal J}_{11}^0\,{\cal J}_1^- + {\cal J}_{22}^0\,{\cal J}_2^-
      + {\cal J}_{33}^0\,{\cal J}_3^-  \,)
      + 6 \,({\cal J}_1^- + {\cal J}_2^- + {\cal J}_3^-)\ +
\end{equation}
\[
2\,\bigg({\cal J}_{11}^0\,({\cal J}_2^- + {\cal J}_3^- ) +  {\cal J}_{22}^0\,({\cal J}_1^- +
      {\cal J}_3^-  ) +   {\cal J}_{33}^0\,({\cal J}_1^- + {\cal J}_2^-  ) -
      {\cal J}_{31}^0\,{\cal J}_2^- - {\cal J}_{23}^0\,{\cal J}_1^- - {\cal J}_{12}^0\,{\cal J}_3^- \bigg)
      \ ,
\]
and
\begin{equation}
     L_1 \ = \  {\cal J}_{21}^0\,-\,{\cal J}_{31}^0\, + \,{\cal J}_{32}^0\,-{\cal J}_{12}^0\, +
     \, {\cal J}_{13}^0\,-\,{\cal J}_{23}^0\, \ .
\label{integral-J}
\end{equation}

The remarkable property of the algebraic operator ${\De_R}(\rho_{ij})$ (\ref{addition3-3rho}) is its gauge-equivalence to the Schr\"odinger operator.
Making the gauge transformation with determinant (\ref{gmn33-rho-det}), (\ref{gmn33-rho-det-gamma}) as the factor,
\[
\Gamma \ = \ D^{-1/4}\ \sim \ \frac{1}{{\ta_1^{1/4}(4 \ta_2-\ta_1^2)}^{1/4}}  \ ,
\]
see also (\ref{gammas}), we find that
\begin{equation}
         \Gamma^{-1}\, {\De_R}(\rho_{ij})\,\Gamma \ =
        \  \De_{LB}(\rho_{ij}) - \tilde V \ ,
\label{HLB3}
\end{equation}
where the effective potential
\[
{\tilde V}(\rho_{ij}) \ =\ \frac{9 }{8 \left(\rho _{12}+\rho _{13}+\rho _{23}\right)}\ + \
  \frac{\left(\rho _{12}+\rho _{13}+\rho _{23}\right)}
  {2\left(\rho _{12}^2+\rho _{13}^2+\rho _{23}^2 -2 \rho _{12} \rho _{13}-
                     2 \rho _{12} \rho _{23}-2 \rho _{13} \rho _{23}\right)}\ .
\]
Note that in $r$-coordinates
\[ \frac{4}{(r_{12}+r_{13}-r_{23})(r_{12}+r_{23}-r_{13})(r_{13}+r_{23}-r_{12})}\ =\
\]
\[\frac{1}{r_{12}r_{13}r_{23}}\left(\frac{r_{23}}{r_{12}+r_{13}-r_{23}}+\frac{r_{13}}
{r_{12}+r_{23}-r_{13}}+\frac{r_{12}}{r_{13}+r_{23}-r_{12}}+1\right),
\]
and
\[
\frac{1}{(r_{23}+r_{13}-r_{12})(r_{23}+r_{12}-r_{13})(r_{13}+r_{12}-r_{23})(r_{12}+r_{13}+r_{23})}
\]
\[=\frac{1}{8 r_{23}r_{13}(r_{23}+r_{13})}\left[ \frac{1}{r_{23}+r_{13}-r_{12}}+ \frac{1}{r_{23}+r_{13}+r_{12}}\right]
\]
\[
 + \frac{1}{8 r_{23} r_{13}r_{12}}\left[\frac{1}{r_{12}-r_{13}+r_{23}}+\frac{1}{r_{12}+r_{13}-r_{23}}\right]\ ,
\]
thus, the effective potential can be written differently,
\[
 {\tilde V}(r_{ij}) \ =\ \frac{9}{8\left(r_{12}^2+r_{13}^2+r_{23}^2\right)}
\]
\[	
+\ \frac{r_{12}^2+r_{13}^2+r_{23}^2}{16}
\left[ \frac{1}{r_{13}r_{23}(r_{13}+r_{23})}\left(\frac{1}{r_{13}+r_{23}-r_{12}}+\frac{1}{r_{12}+r_{13}+r_{23}}\right)
\right.
\]
\[
\left. +\ \frac{1}{r_{12}r_{13}r_{23}}\left(\frac{1}{r_{12}+r_{23}-r_{13}}+\frac{1}{r_{12}+r_{13}-r_{23}}\right)\right]\ .
\]

In turn,
\[
  \De_{LB}(\rho_{ij}) \ =\ 4(\rho_{12} \pa^2_{\rho_{12}} + \rho_{13} \pa^2_{\rho_{13}} +\rho_{23} \pa^2_{\rho_{23}})
\]
\[
   + \ 2\, \bigg((\rho_{12} + \rho_{13} - \rho_{23})\pa_{\rho_{12}}\pa_{\rho_{13}}\ +
          (\rho_{12} + \rho_{23} - \rho_{13})\pa_{\rho_{12}}\pa_{\rho_{23}}\ +
          (\rho_{13} + \rho_{23} - \rho_{12})\pa_{\rho_{13}}\pa_{\rho_{23}}
  \bigg)
\]
\begin{equation}
\label{LB3}
- 3\, \bigg(\frac{\rho_{12}\pa_{\rho_{12}}+\rho_{13}\pa_{\rho_{13}}+\rho_{23}\pa_{\rho_{23}}}
{\rho _{12}+\rho _{13}+\rho _{23}} \bigg) + 4\, (\pa_{\rho_{12}}+\pa_{\rho_{23}}+ \pa_{\rho_{13}})\ ,
\end{equation}
is the Laplace-Beltrami operator,
\[
   \De_{LB}(\rho_{ij})\ =\ \sqrt D\, \pa_{\mu} \frac{1}{\sqrt D}\, g^{\mu \nu} \pa_{\nu}\ ,\quad \pa_{\nu}\equiv \frac{\pa}{\pa_{\rho_{\nu}}}\ ,
\]
see (\ref{gmn33-rho}), (\ref{gmn33-rho-det}). Eventually, taking into account (\ref{HLB3}) we arrive at the Hamiltonian
\begin{equation}
\label{H-3-3r-r}
    {\cal H}_{rd} (r_{ij}) \ =\ -\De_{LB}(r_{ij}) + \tilde V (r_{ij}) +  V(r_{12},\,r_{13},\,r_{23})
    \ ,
\end{equation}
in the space of relative distances, or
\begin{equation}
\label{H-3-3r-rho}
    {\cal H}_{rd} (\rho_{ij}) \ =\ -\De_{LB}(\rho_{ij}) + \tilde V (\rho_{ij}) +  V(\rho_{ij})\ ,
\end{equation}
in $\rho$-space, see (\ref{rho}). The Hamiltonian (\ref{H-3-3r-r}), or (\ref{H-3-3r-rho}) describes the three-dimensional quantum particle
moving in the curved space with metric $g^{\mu \nu}$. The Ricci scalar, see e.g. \cite{Eis}, for this space is equal to
\[ Rs \ = \ -\frac{41\,{(\rho_{12}+\rho_{13}+\rho_{23})}^2  -84\,(\rho_{12}\,\rho_{13}+\rho_{12}\,\rho_{23}+\rho_{23}\,\rho_{13})}
   {12\,(\rho_{12}+\rho_{13}+\rho_{23})\left({(\rho_{12}+\rho_{13}+\rho_{23})}^2-4\,(\rho_{12}\,\rho_{13}+\rho_{12}\,\rho_{23}
   +\rho_{23}\,\rho_{13})\right)}
\]
\[
\ = \ \frac{-84\,\tau_2\ +\ 41\,{\tau_1}^2   }{12\,\tau_1(4\,\tau_2\ -\ \tau_1^2)}\ .
\]
It is singular at the boundary of the configuration space. The Cotton tensor, see e.g. \cite{Eis}, for this metric is nonzero, so the space is not conformally flat.

Making the de-quantization of (\ref{H-3-3r-rho}) we arrive at a three-dimensional classical system which is characterized by the Hamiltonian,
\begin{equation}
\label{H-3-3r-rho-class}
    {\cal H}_{rd}^{(c)} (\rho_{ij}) \ =\ g^{\mu \nu}(\rho_{ij})\,P_{\mu}\, P_{\nu} + \tilde V (\rho_{ij}) +  V(\rho_{ij})\ ,
\end{equation}
where $P_i\, P_j, \, i,j=1,2,3$ are classical momenta in $\rho$-space and $g^{\mu \nu}(\rho_{ij})$ is given by (\ref{gmn33-rho}).
Here the underlying manifold (zero-potential case) admits an $so(3)$ algebra
of constants of the motion linear in the momenta, i.e., Killing vectors. Thus, the free Hamilton-Jacobi equation is integrable.
However, it admits no separable coordinate system.

\subsection{{(Quasi)}-exact-solvability}

Let us take the function
\begin{equation}
\label{psi_cal-r-d3}
       \Psi_0(\rho_{12},\,\rho _{13},\,\rho _{23}) \ = \ \ta_1^{1/4}
{(4\,\ta_2-\tau_1^2)}^{\frac{\gamma}{2}}\,e^{-\om\,\ta_1 - \frac{A}{2}\,\ta_1^2} \ ,
\end{equation}
where $\gamma,\,\om > 0$ and $A \geq 0$ are constants and $\ta$'s are given by (\ref{gammas}),
and seek the potential for which this (\ref{psi_cal-r-d3}) is the
ground state function for the Hamiltonian ${\cal H}_r(\rho_{ij})$, see (\ref{H-3-3r-rho}). This potential can be found immediately by calculating the ratio
$$\frac{\De_{LB}(\rho_{ij}) \Psi_0 }{ \Psi_0}\ =\ V_0 - E_0 \,.$$
The  result is
\[
  V_0(\ta_1,\,\ta_2)\ = \ \frac{9}{8 \ta_1} + \,\gamma(\gamma-1)
  \left ( \frac{2\ta_1}{4\ta_2-\tau_1^2} \right )\ +
\]
\begin{equation}
 6\,\om^2\,\ta_1\, +\, 6\,A\,\ta_1\,(2\,\om\,\ta_1\, -\, 2\gamma - 3)\,
   +\, 6\, A^2\ta_1^3      \ ,
\label{VQES-0}
\end{equation}
with the energy of the ground state
\begin{equation}
  E_0\ =\ 12\,\om\,(1+\,\gamma) \ .
\label{EQES-0}
\end{equation}

Now, let us take the Hamiltonian ${\cal H}_{rd,0} \equiv  -\De_{LB} + V_0$, see (\ref{H-3-3r-rho}), with potential (\ref{VQES-0}), subtract $E_0$ (\ref{EQES-0}) and make the gauge
rotation with $\Psi_0$ (\ref{psi_cal-r-d3}). As the result we obtain the $sl(4, {\bf R})$-Lie-algebraic operator with additional potential $\De V_N$, \cite{Turbiner:1988,Turbiner:2016}
\[
   \Psi_0^{-1}\,(-{\De_{LB}} + V_0 - E_0)\,\Psi_0\ =\ -{\De_R}({\cal J})\
   +\ 2(1-2\,\gamma)\,({\cal J}_1^- + {\cal J}_2^- + {\cal J}_3^-)\ +
\]
\begin{equation}
     12\,\om\,({\cal J}_{11}^0  +{\cal J}_{22}^0 + {\cal J}_{33}^0) + 12\,A\,\left({\cal J}_1^+(N) + {\cal J}_2^+(N)  + {\cal J}_3^+(N) \right)\ +\ \De V_N
\label{HQES-0-Lie}
\end{equation}
\[
  \equiv h^{(qes)}(J)\ +\ \De V_N\ ,
\]
see (\ref{HRex}), where
\[
    \De V_N\ =\ 12 \,A\,N \ta_1\ .
\]
It is evident that for integer $N$ the operator $h(J)$ has a finite-dimensional invariant subspace
${\cal P}^{(3)}_{N}$, (\ref{P3}), with $\dim {\cal P}^{(3)}_{N} \sim N^3$ at large $N$.
Finally, we arrive at the quasi-exactly-solvable Hamiltonian in the space of relative distances:
\begin{equation}
\label{HQES-3-3}
    {\cal H}_{rd,qes}(\rho_{ij}) \ =\ -\De_{LB}(\rho_{ij}) + V_N^{(qes)}(\rho_{ij})\ ,
\end{equation}
cf.(\ref{Hrel-final}), where
\[
   V^{(qes,N)}(\ta_1,\,\ta_2)\ = \ \frac{9}{8 \ta_1} + \,\gamma(\gamma-1)\left ( \frac{2\ta_1}{4\ta_2-\ta_1^2}\right )\ +
\]
\begin{equation}
  +\ 6\,\om^2\,\ta_1\
  +\ 6\,A\,\ta_1\,(2\,\om\,\ta_1\,-\,2\gamma\ -\ 2 N\, -\, 3)\ +\ 6\, A^2\ta_1^3 \ .
\label{VQES-N}
\end{equation}
For this potential $\sim N^3$ eigenstates can be found by algebraic means. They have the factorized form of the polynomial multiplied by $\Psi_0$ (\ref{psi_cal-r-d3}),
\[
   \mbox{Pol}_N (\rho_{12}, \rho_{13},\rho_{23})\ \Psi_0 (\ta_1, \ta_2)\ .
\]
(Note that for given $N$ we can always choose appropriate values of $\gamma$ such that the boundary terms vanish for polynomials in the invariant subspace vanish and the Hamiltonian (\ref{HQES-3-3}) acts as a self-adjoint operator.)
These polynomials are the eigenfunctions of the quasi-exactly-solvable algebraic operator
\begin{equation}
\label{hQES-N-alg}
  h^{(qes)}(\rho) \ = \
\end{equation}
\[
   -4(\rho_{12} \pa^2_{\rho_{12}} + \rho_{13} \pa^2_{\rho_{13}} +\rho_{23} \pa^2_{\rho_{23}})
\]
\[
 -2\left( (\rho_{12} + \rho_{13} - \rho_{23})\pa_{\rho_{12}}\pa_{\rho_{13}}
          +(\rho_{12} + \rho_{23} - \rho_{13})\pa_{\rho_{12}}\pa_{\rho_{23}}
          +(\rho_{13} + \rho_{23} - \rho_{12})\pa_{\rho_{13}}\pa_{\rho_{23}}\right)
\]
\[
   + 2(1-2\,\gamma)(\pa_{\rho_{12}} + \pa_{\rho_{13}}+ \pa_{\rho_{23}}) + 12\,\om(\rho_{12}\pa_{\rho_{12}}+\rho_{13}\pa_{\rho_{13}}+\rho_{23}\pa_{\rho_{23}})
\]
\[
-12\,A\,(\rho_{12}+ \rho_{13}+\rho_{23})(\rho_{12}\,\pa_{\rho_{12}} + \rho_{13}\,\pa_{\rho_{13}}+ \rho_{23}\,\pa_{\rho_{23}}-N   )
\]
which is the quasi-exactly-solvable $sl(4,\,{\bf R})$-Lie-algebraic operator
\begin{equation}
h^{(qes)}(J) \ = \  -4(\, {\cal J}_{11}^0\,{\cal J}_1^- + {\cal J}_{22}^0\,{\cal J}_2^-
      + {\cal J}_{33}^0\,{\cal J}_3^-  \,)
\label{hQES-N-Lie}
\end{equation}
\[
 - 2\,\bigg({\cal J}_{11}^0\,({\cal J}_2^- + {\cal J}_3^- ) +  {\cal J}_{22}^0\,({\cal J}_1^- +
      {\cal J}_3^-  ) +   {\cal J}_{33}^0\,({\cal J}_1^- + {\cal J}_2^-  ) -
      {\cal J}_{31}^0\,{\cal J}_2^- - {\cal J}_{23}^0\,{\cal J}_1^- - {\cal J}_{12}^0\,{\cal J}_3^- \bigg)
\]
\[
+2\,(1-2\,\gamma)\,( {\cal J}_1^- + {\cal J}_2^- + {\cal J}_3^-  ) + 12\,\om\,({\cal J}_{11}^0 + {\cal J}_{22}^0 + {\cal J}_{33}^0)
\]
\[
+ 12\,A\,(\, J_1^+(N)+J_2^+(N)+J_3^+(N)\, )  \ ,
\]
cf. (\ref{HQES-0-Lie}).

As for the original problem (\ref{Hrel-Mod}) in the space of relative motion
\[
   {\tilde {\cal H}}_R\,\Psi(r_{ij}) \equiv \ \bigg(- {\De_R}(r_{ij}) + V(r_{ij})\bigg)\, \Psi(r_{ij})\ =\ E\Psi(r_{ij})\ ,\ \Psi \in L_2 ({\bf \tilde R})\ ,
\]
the potential for which quasi-exactly-solvable, polynomial solutions occur of the form
\[
   \mbox{Pol}_N (\rho_{12}, \rho_{13},\rho_{23})\ \Gamma \  \Psi_0 (\ta_1, \ta_2) \ ,
\]
where $\Gamma \sim   D^{-1/4}$, see (\ref{gmn33-rho-det-gamma}), is given by
\[
V_{relative}^{(qes,N)}(\ta) \ = \  {\bigg(\gamma-\frac{1}{2}\bigg)}^2\left (\frac{2\ta_1}{4\ta_2-\ta_1^2}\right )\ +
\]
\begin{equation}
  +\ 6\,\om^2\,\ta_1\
  +\ 6\,A\,\ta_1\,(2\,\om\,\ta_1\,-\,2\gamma\ -\ 2 N\, -\, 3)\
   +\ 6\, A^2\ta_1^3      \ ,
\label{VQES-N-rel}
\end{equation}
cf. (\ref{VQES-N}); it does not depend on $\ta_3$.

If the parameter $A$ vanishes in (\ref{psi_cal-r-d3}), (\ref{VQES-N}) and (\ref{HQES-0-Lie}), (\ref{hQES-N-Lie}) we will arrive at the
exactly-solvable problem, where $\Psi_0$ (\ref{psi_cal-r-d3}) at $A=0$, plays the role of the ground state function,
\begin{equation}
\label{psi_cal-r-d2exact}
   \Psi_0(\rho_{12},\,\rho _{13},\,\rho _{23}) \ = \ \ta_1^{1/4}
    {(4\,\ta_2-\ta_1^2)}^{\frac{\gamma}{2}}\,e^{-\om\,\ta_1} \ ,
\end{equation}
which does not depend on $\ta_3$. In this case the $sl(4, {\bf R})$-Lie-algebraic operator (\ref{hQES-N-Lie}) contains no raising generators $\{{\cal J}^+(N)\}$ and becomes
\[
  h^{(exact)} = -{\De_R}({\cal J}) + 2(1-2\,\gamma)\,({\cal J}_1^- + {\cal J}_2^- + {\cal J}_3^-) +
     12\,\om\,({\cal J}_{11}^0  +{\cal J}_{22}^0 + {\cal J}_{33}^0)\ ,
\]
see (\ref{HRex}), and, hence, preserves the infinite flag of finite-dimensional invariant subspaces
${\cal P}^{(3)}_{N}$ (\ref{P3}) at $N=0,1,2\ldots$\,. The potential (\ref{VQES-N}) becomes
\begin{equation}
   V^{(es)}(\ta_1,\,\ta_2)\ = \ \frac{9}{8 \ta_1} + \,\gamma(\gamma-1)\left ( \frac{2\ta_1}{4\ta_2-\ta_1^2}\right )\ +
  \ 6\,\om^2\,\ta_1\ =\
\label{VES}
\end{equation}
\[
   =\  \frac{9}{8 \left(\rho _{12}+\rho _{13}+\rho _{23}\right)}\ +\ 6\om^2 \left(\rho _{12}+\rho _{13}+
       \rho _{23}\right)
\]
\[
  -\ \gamma(\gamma-1)
\left ( \frac{2(\rho_{12}+\rho_{13}+\rho_{13})}{\rho_{12}^2+\rho_{13}^2+\rho_{23}^2-2\rho_{12}\rho_{13}-2\rho_{12}\rho_{23}
-2\rho_{13}\rho_{23}}
\right )\
\]

\[
   =\ \frac{9}{8 \left(r_{12}^2+r_{13}^2+r_{23}^2\right)}\ +\ 6\om^2 \left(r_{12}^2+r_{13}^2+r_{23}^2\right)
\]
%
\[
  +\ \gamma(\gamma-1)\frac{r_{12}^2+r_{13}^2+r_{23}^2}{16}\left[
   \frac{1}{r_{13}r_{23}(r_{13}+r_{23})}\left(\frac{1}{r_{13}+r_{23}
-r_{12}}+\frac{1}{r_{12}+r_{13}+r_{23}}\right)\right.\]
\[\left.+\frac{1}{r_{12}r_{13}r_{23}}\left(\frac{1}{r_{12}+r_{23}-r_{13}}+\frac{1}{r_{12}+r_{13}-r_{23}}\right)\right].
\]

Eventually, we arrive at the exactly-solvable Hamiltonian in the space of relative distances
\begin{equation}
\label{HES-3-2}
    {\cal H}_{rd,es}(\rho_{ij}) \ =\ -\De_{LB}(\rho_{ij}) + V^{(es)}(\rho_{ij})\ ,
\end{equation}
where the spectra of energies
\[
    E_{n_1, n_2, n_3}\ =\  12 \om (n_1 + n_2 + n_3 + \gamma + 1)\ ,\quad n_1, n_2, n_3=0,1,2,\ldots
\]
is equidistant. All eigenfunctions have the factorized form of
a polynomial multiplied by $\Psi_0$ (\ref{psi_cal-r-d2exact}),
\[
   \mbox{Pol}_N (\rho_{12}, \rho_{13},\rho_{23})\ \Psi_0 (\ta_1, \ta_2)\ ,\quad N=0,1,\ldots\ .
\]

These polynomials are eigenfunctions of the exactly-solvable algebraic operator
\[
    h^{(exact)}(\rho)\ =  \ -4(\rho_{12} \pa^2_{\rho_{12}} + \rho_{13} \pa^2_{\rho_{13}} +\rho_{23} \pa^2_{\rho_{23}}) + {(2-4\,\gamma)}(\pa_{\rho_{12}} + \pa_{\rho_{13}}+ \pa_{\rho_{23}}) + 12\,\om(\rho_{12}\pa_{\rho_{12}}+\rho_{13}\pa_{\rho_{13}}+\rho_{23}\pa_{\rho_{23}})
\]
\begin{equation}
\label{hES-rho}
  -2 (\rho_{12} + \rho_{13} - \rho_{23})\pa_{\rho_{12}}\pa_{\rho_{13}}
          -2(\rho_{12} + \rho_{23} - \rho_{13})\pa_{\rho_{12}}\pa_{\rho_{23}}
          -2(\rho_{13} + \rho_{23} - \rho_{12})\pa_{\rho_{13}}\pa_{\rho_{23}} \ ,
\end{equation}
or, equivalently, of the exactly-solvable $sl(4,{\bf R})$-Lie-algebraic operator
\[
     h^{(exact)}(J) \ = \  -4(\, {\cal J}_{11}^0\,{\cal J}_1^- + {\cal J}_{22}^0\,{\cal J}_2^-
      + {\cal J}_{33}^0\,{\cal J}_3^-  \,)
\]
\[
  - 2\,\bigg({\cal J}_{11}^0\,({\cal J}_2^- + {\cal J}_3^- ) +  {\cal J}_{22}^0\,({\cal J}_1^- +
      {\cal J}_3^-  ) +   {\cal J}_{33}^0\,({\cal J}_1^- + {\cal J}_2^-  ) -
      {\cal J}_{31}^0\,{\cal J}_2^- - {\cal J}_{23}^0\,{\cal J}_1^- - {\cal J}_{12}^0\,{\cal J}_3^- \bigg)
\]
\begin{equation}
\label{hES-N-Lie}
+2\,(1-2\,\gamma)\,( {\cal J}_1^- + {\cal J}_2^- + {\cal J}_3^-  ) + 12\,\om\,({\cal J}_{11}^0 + {\cal J}_{22}^0 + {\cal J}_{33}^0)\ .
\end{equation}
Those polynomials are orthogonal w.r.t. $\Psi_0^2$\,,  (\ref{psi_cal-r-d3}) at $A=0$, their domain is given by (\ref{CFrho}).
Being written in variables $w_{1,2,3}$, see above, they are factorizable, $F(w_1,w_2)\, f(w_3)$.
To the best of our knowledge these orthogonal polynomials have not been studied in literature.

The Hamiltonian with potential (\ref{VES}) can be considered as a three-dimensional generalization of the 3-body Calogero model \cite{Calogero:1969},
see also \cite{RT:1995}, \cite{ST:2015}, with loss of the
property of pairwise interaction. Now the potential of interaction contains two- and three-body interaction terms. If $\gamma=0,1$ in (\ref{VES})
we arrive at the celebrated harmonic oscillator potential in the space of relative distances, see e.g. \cite{Green}. In turn, in the space of
relative motion this potential contains no singular terms and becomes,
\[
     V \ =\ 6\om^2 \ta_1\ =\ 6\om^2(\rho_{12} + \rho_{13} + \rho_{23})\ =\ 6\om^2(r^2_{12} + \ r^2_{13} + \ r^2_{23})\ ,
\]
see \cite{Green}.

The quasi-exactly-solvable $sl(4,\,{\bf R})$-Lie-algebraic operator $h^{(qes)}(J)$\,, (\ref{hQES-N-Lie})
as well as the exactly-solvable operator as a degeneration at $A=0$, written originally in
$\rho$ variables (\ref{hQES-N-alg}) can be rewritten in $\ta$ variables (\ref{gammas}).
Surprisingly, this operator is algebraic (!) as well
\begin{equation}
\label{hQES-N-tau}
    h^{(qes)}(\tau) \ = \ -6\,\ta_1\pa_1^2 -2\ta_1(7\ta_2-\ta_1^2)\pa_2^2 -2\ta_3(6\ta_2-\ta_1^2)\pa_3^2
   -\,24\,\ta_2\pa_{1,2}^2 - 36\ta_3\pa_{1,3}^2\ -
\end{equation}
\[
 2\,(4\ta_2^2+9\ta_1\ta_3-\ta_1^2\ta_2)\pa_{2,3}^2 -18\pa_1 -14\ta_1\pa_2-2(7\ta_2-\ta_1^2)\pa_3\ +
\]
\[
 2(1-2\gamma)(3\pa_1+2\ta_1\pa_2+\ta_2\pa_3) +12\om\,(\ta_1\pa_1+2\ta_2\pa_2+3\ta_3\pa_3)\ +
\]
\[
  12\,A\ta_1(\ta_1\pa_1 + 2\ta_2\pa_2 + 3\ta_3\pa_3 - N) \ .
\]
Evidently, it remains algebraic at $A=0$,
\begin{equation}
\label{hES-N-tau}
    h^{(es)}(\tau) \ = \ -6\,\ta_1\pa_1^2 -2\ta_1(7\ta_2-\ta_1^2)\pa_2^2 -2\ta_3(6\ta_2-\ta_1^2)\pa_3^2
   -\,24\,\ta_2\pa_{1,2}^2 - 36\ta_3\pa_{1,3}^2\ -
\end{equation}
\[
 2\,(4\ta_2^2+9\ta_1\ta_3-\ta_1^2\ta_2)\pa_{2,3}^2 -18\pa_1 -14\ta_1\pa_2-2(7\ta_2-\ta_1^2)\pa_3\ +
\]
\[
 2(1-2\gamma)(3\pa_1+2\ta_1\pa_2+\ta_2\pa_3) +12\om\,(\ta_1\pa_1+2\ta_2\pa_2+3\ta_3\pa_3)\ ,
\]
becoming the exactly-solvable one. Note that the (quasi)-exactly-solvable operator (\ref{hQES-N-tau}) (and (\ref{hES-N-tau})) admits the integral
\[
     -L_1^2 \ =\
     \left(27\tau_3^2 + 4\tau_3\tau_1^3 - 18\tau_3\tau_2\tau_1 - \ta_2^2 \ta_1^2 + 4\tau_2^3 \right) \pa^2_{\tau_3}\ +\ \left(27\tau_3 + 2\tau_1^3 - 9\tau_1\tau_2 \right) \pa_{\tau_3}\ ,
\]
cf. (\ref{integral}), $[h^{(qes)}(\tau), L_1^2]=0$. It involves derivatives w.r.t. $\ta_3$ only.

It can be immediately checked that the quasi-exactly-solvable operator (\ref{hQES-N-tau}) has the finite-dimensional invariant subspace in polynomials,
\begin{equation}
\label{P3-tau}
     {\mathcal P}^{(1,2,3)}_{N}\ =\ \langle \ta_1^{p_1} \ta_2^{p_2} \ta_3^{p_3} \vert \
     0 \le p_1+2p_2+3p_3 \le N \rangle\ ,
\end{equation}
cf. (\ref{P3}). This finite-dimensional space appears as a finite-dimensional representation space
of the algebra of differential operators $h^{(3)}$ which was discovered in the relation with $H_3$ (non-crystallographic) rational Calogero model \cite{GT} as its hidden algebra.

The algebra $h^{(3)}$ is infinite-dimensional but finitely-generated, for discussion see \cite{GT}. Their generating elements can be split into two classes. The first class of generators (lowering and Cartan
operators) act in $\mathcal{P}^{(1,2,3)}_N$ for any $N$ and
therefore they preserve the flag $\mathcal{P}^{(1,2,3)}$. The second
class operators (raising operators) act on the space
$\mathcal{P}^{(1,2,3)}_N$ only.

Let us introduce the following notation for the derivatives:
\[
\pa_i\equiv\frac{\pa}{\pa\tau_i}\ ,\quad
\pa_{ij}\equiv\frac{\pa^2}{\pa\tau_{i}\pa\tau_{j}}\
,\quad\pa_{ijk}\equiv\frac{\pa^3}{\pa\tau_{i}\pa\tau_{j}\pa\tau_{k}}\
.
\]
The first class of generating elements consist of the 22 generators where 13 of them are the first order operators
\begin{equation}
\begin{aligned}
\label{ops_1}
& T_0^{(1)}=\pa_1\,, && T_0^{(2)}=\pa_2\,, && T_0^{(3)}=\pa_3\,,\\
& T_1^{(1)}=\tau_1\pa_1\,, && T_2^{(2)}=\tau_2\pa_2\,, && T_3^{(3)}=\tau_3\pa_3\,,\\
& T_1^{(3)}=\tau_1\pa_3\,, && T_{11}^{(3)}=\tau_1^2\pa_3\,, && T_{111}^{(3)}=\tau_1^3\pa_3\,,\\
& T_1^{(2)}=\tau_1\pa_2\,, && T_{11}^{(2)}=\tau_1^2\pa_2\,, && T_2^{(3)}=\tau_2\pa_3\,,\\
& &&T_{12}^{(3)}=\tau_1\tau_2\pa_3\ ,&&
\end{aligned}
\end{equation}
the 6 are of the second order
\begin{equation}
\begin{aligned}
\label{ops_2}
& T_2^{(11)}=\tau_2\pa_{11}\,, && T_{22}^{(13)}=\tau_2^2\pa_{13}\,, && T_{222}^{(33)}=\tau_2^3\pa_{33}\,,\\
& T_3^{(12)}=\tau_3\pa_{12}\,, && T_3^{(22)}=\tau_3\pa_{22}\,, &&
T_{13}^{(22)}=\tau_1\tau_3\pa_{22}\ ,
\end{aligned}
\end{equation}
and 2 are of the third order
\begin{equation}
\begin{aligned}
\label{ops_3}
& T_3^{(111)}=\tau_3\pa_{111}\,, &&
T_{33}^{(222)}=\tau_3^2\pa_{222}\ .
\end{aligned}
\end{equation}

The generators of the second class consist of 8 operators where 1 of them is of the first order
\begin{equation}
\label{R1}
T_1^+ = \ta_1 T_0\ ,
\end{equation}
4 are of the second order
\begin{equation}
\begin{aligned}
\label{R2}
& T_{2,-1}^+=\tau_2\pa_1T_0\,, &&
T_{3,-2}^+=\tau_3\pa_2T_0\,,  && T_{22,-3}^+ = \ta_2^2\pa_3T_0\,,
&& T_2^+ = \tau_2T_0(T_0+1)\ ,
\end{aligned}
\end{equation}
and 3 are of the third order
\begin{equation}
\begin{aligned}
\label{R3}
& T_{3,-11}^{+}=\tau_3\pa_{11}T_0\ , &&
T_{3,-1}^+=\tau_3\pa_1T_0(T_0+1)\ , &&
T_3^+=\tau_3T_0(T_0+1)(T_0+2)\ ,
\end{aligned}
\end{equation}
where we have introduced the diagonal operator (the Euler-Cartan generator)
\begin{equation}
\label{jo}
T_0=\tau_1\pa_1+2\tau_2\pa_2+3\tau_3\pa_3 - N\ .
\end{equation}
for a convenience. In fact, this operator is the identity operator,
it is of the zeroth order and, hence, it belongs to the first class.

It is not surprising that the algebraic operator $h^{(qes)}(\tau)$ (\ref{hQES-N-tau}) can be rewritten in terms of generators of the $h^{(3)}$-algebra,
\begin{equation}
\begin{aligned}
   h^{(qes)}(T) \ = & \  - \bigg[6\,T_1^{(1)}\,T_0^{(1)} + 2\,(7\,T_2^{(2)} - T_{11}^{(2)})\,T_1^{(2)} + T_3^{(3)}(6\,T_2^{(3)} - T_{11}^{(3)})
\\ &  \ + T_0^{(1)}\,(24\,T_2^{(2)} +36\,T_3^{(3)}) + 2\,(4\,T_2^{(3)}\,T_2^{(2)}+9\,T_1^{(2)}\,T_3^{(3)}-\,T_{11}^{(3)}\,T_2^{(2)})
\\ &  + 2\,(9\,T_0^{(1)}+7\,T_1^{(2)}) + 2\,(7\,T_2^{(3)}-T_{11}^{(3)})
                          \bigg]
\label{hj}
\end{aligned}
\end{equation}
\[
+2\,(1-2\,\gamma)\,(T_2^{(3)} + 2\,T_1^{(2)} + 3\,T_0^{(1)}   ) + 12\,\om\,(J_0+N) + 12\,A\,J_1^+\ ,
\]
as well as the algebraic operator $h^{(es)}(\tau)$ (\ref{hES-N-tau}), which occurs at $A=0$, can be rewritten in terms of generators of the $h^{(3)}$-algebra,
\begin{equation}
\begin{aligned}
   h^{(es)}(T) \ = & \  -\bigg[6\,T_1^{(1)}\,T_0^{(1)}  +2\,(7\,T_2^{(2)} - T_{11}^{(2)})\,T_1^{(2)} + T_3^{(3)}(6\,T_2^{(3)} - T_{11}^{(3)})
\\ &  \ + T_0^{(1)}\,(24\,T_2^{(2)} +36\,T_3^{(3)}) + 2\,(4\,T_2^{(3)}\,T_2^{(2)}+9\,T_1^{(2)}\,T_3^{(3)}-\,T_{11}^{(3)}\,T_2^{(2)})
\\ &  + 2\,(9\,T_0^{(1)}+7\,T_1^{(2)}) + 2\,(7\,T_2^{(3)}-T_{11}^{(3)})
                        \bigg]
\label{hj1}
\end{aligned}
\end{equation}
\[
+2\,(1-2\,\gamma)\,(T_2^{(3)} + 2\,T_1^{(2)} + 3\,T_0^{(1)}   ) + 12\,\om\,J_0\ ,
\]
where without a loss of generality we put $N=0$.

It can be immediately verified that with respect to the action of the operator (\ref{hQES-N-tau}) the
finite-dimensional invariant subspace (\ref{P3-tau}) is reducible: it preserves
\begin{equation}
\label{P2-tau}
     {\mathcal P}^{(1,2)}_{N}\ \equiv \ \langle \ta_1^{p_1} \ta_2^{p_2} \vert \
     0 \le p_1+2p_2 \le N \rangle\ \subset {\mathcal P}^{(1,2,3)}_{N}\ .
\end{equation}
The operator which acts on ${\mathcal P}^{(1,2)}_{N}$ has the form,
\begin{equation}
\label{hQES-N-tau-2}
    h^{(qes)}(\ta_1,\ta_2) \ = \ -6\,\ta_1\pa_1^2 -2\ta_1(7\ta_2-\ta_1^2)\pa_2^2 -\,24\,\ta_2\pa_{1,2}^2\ -\
    12(1+\gamma)\pa_1 - 2(5+4\gamma)\ta_1\pa_2\
\end{equation}
\[
 +\ 12\om\,(\ta_1\pa_1+2\ta_2\pa_2)\ +\ 12\,A\ta_1(\ta_1\pa_1 + 2\ta_2\pa_2 - N) \ ,
\]
cf. (\ref{eq-psi}).
It has $\sim {N}^2$ polynomial eigenfunctions which depends on two variables $\ta_{1,2}$ only. Note that the space ${\mathcal P}^{(1,2)}_{N}$ is finite-dimensional representation space of the non-semi-simple Lie algebra $gl(2,{\bf R}) \oplus R^3$ realized by the first order differential operators,
\cite{gko1} (see also \cite{Turbiner:1994}, \cite{ghko}, \cite{Turbiner:1998}),
\[ t_1\  =\  \pa_{\ta_1} \ , \]
\[ t_2 ({ N})\  =\ {\ta_1} \pa_{\ta_1}\ -\ \frac{{ N}}{3} \ ,
 \ t_3 ({ N})\  =\ 2 {\ta_2}\pa_{\ta_2}\ -\ \frac{{ N}}{3}\ ,\]
\[ t_4 ({ N})\  =\ {\ta_1}^2 \pa_{\ta_1} \  +\ 2 {\ta_1} {\ta_2} \pa_{\ta_2} \ - \ { N} {\ta_1} \ ,\]
\begin{equation}
\label{gr}
 r_{i}\  = \ {\ta_1}^{i}\pa_{\ta_2}\ ,\quad i=0, 1, 2\ .
\end{equation}
The operator (\ref{hQES-N-tau-2}) can be rewritten in terms of $gl(2,{\bf R}) \oplus R^3$ operators
\[
     h^{(qes)}(t,r)\ =\ -6\,r_1 t_1 - 14 (t_3 + \frac{{ N}}{3}) r_1 + 2 r_2 r_1 -  24 t_1 (t_3 + \frac{{ N}}{3})
\]
\[
     - 12(1+\gamma)t_1 - 2(5+4\gamma) r_1 + 12\,\om (t_2 + t_3 + N)  +12 A t_4\ .
\]

The space (\ref{P3-tau}) is reducible further: the operator (\ref{hQES-N-tau}) (and also the operator (\ref{hQES-N-tau-2})) preserves
\begin{equation}
\label{P1-tau}
     {\mathcal P}^{(1)}_{N}\ \equiv \ \langle \ta_1^{p_1} \vert \
     0 \le p_1 \le N \rangle\ \subset {\mathcal P}^{(1,2)}_{N}\  \subset {\mathcal P}^{(1,2,3)}_{N}\ ,
\end{equation}
as well.
The operator, which acts on ${\mathcal P}^{(1)}_{N}$, has the form,
\begin{equation}
\label{hQES-N-tau-1}
    h^{(qes)}(\ta_1) \ = \ -6\,\ta_1\pa_1^2  -\
 12(1+\gamma)\pa_1 + 12\om\,\ta_1\pa_1\ +  12\,A\ta_1(\ta_1\pa_1 - N) \ .
\end{equation}
It can be rewritten in terms of $sl(2, {\bf R})$ algebra generators,
\begin{equation}
\label{sl2}
    {\cal J}^+(N)\ =\ \ta_1^2\pa_{\ta_1} - N \ta_1\ ,\ {\cal J}^0(N)\ =\ 2\ta_1\pa_{\ta_1} - N\ ,\ {\cal J}^-(N)\ =\ \pa_{\ta_1}\ .
\end{equation}

Eventually, it can be stated that among $\sim N^3$ polynomial eigenfunctions in $\ta$'s variables  of the quasi-exactly-solvable operator (\ref{hQES-N-tau}) there are $\sim N^2$ polynomial eigenfunctions of the quasi-exactly-solvable operator (\ref{hQES-N-tau-2}) and $\sim N$ polynomial eigenfunctions of the quasi-exactly-solvable operator (\ref{hQES-N-tau-1}). Similar situation occurs for the exactly-solvable operator (\ref{hES-N-tau}), see (\ref{hQES-N-tau}) at $A=0$, for which there exist infinitely-many polynomial eigenfunctions in $\ta$'variables. Among these eigenfunctions there exists the infinite family of the polynomial eigenfunctions in $\ta_{1,2}$ variables, which are eigensolutions of the operator
\begin{equation}
\label{hES-N-tau-2}
    h^{(es)}(\ta_1,\ta_2) \ = \ -6\,\ta_1\pa_1^2 -2\ta_1(7\ta_2-\ta_1^2)\pa_2^2 -\,24\,\ta_2\pa_{1,2}^2\ -\
    12\pa_1 - 10\ta_1\pa_2\ -\
\end{equation}
\[
 4\gamma(3\pa_1+2\ta_1\pa_2) +12\om\,(\ta_1\pa_1+2\ta_2\pa_2)\ .
\]
Besides that there exists the infinite family of the polynomial eigenfunctions in $\ta_{1}$ variable, which are eigensolutions of the operator
\begin{equation}
\label{hES-N-tau-1}
    h^{(es)}(\ta_1) \ = \ -6\,\ta_1\pa_1^2\ +\ 12(\om\,\ta_1-\gamma -1)\pa_1\ ,
\end{equation}
they are nothing but the Laguerre polynomials.

\section*{Conclusions}

In this paper we found the Schr\"odinger type equation in the space ${\bf \tilde R}$ of relative distances $\{ r_{ij} \}$,
\begin{equation}
\label{H3}
     {\cal H}_{rd} \Psi(r_{12},\,r_{13},\,r_{23}) \ =\ E \Psi(r_{12},\,r_{13},\,r_{23})\ ,\
    {\cal H}_{rd} \ =\ -\De_{LB}(r_{ij}) +  V(r_{12},\,r_{13},\,r_{23})
    \ ,
\end{equation}
where the Laplace-Beltrami operator $\De_{LB}$, see e.g. (\ref{LB3}), makes sense of the  kinetic energy of a three-dimensional particle in curved space with metric (\ref{gmn33-rho}). This equation describes angle-independent solutions of the original 3-body problem in three-dimensional space (\ref{Hgen}),
including the ground state. Hence, finding the ground state involves the solution of the differential equation in three variables, contrary to the original six-dimensional Schr\"odinger equation of the relative motion. Since the Hamiltonian ${\cal H}_r$ is Hermitian, the variational method can be employed with only three-dimensional integrals involved.

The gauge-rotated Laplace-Beltrami operator, with determinant of the metric $D$ raised to a certain degree as the gauge factor, appears as the algebraic operator both in the variables which are squares of relative distances and which are the elementary symmetric polynomials in squares of relative distances as arguments. The former algebraic operator has the  hidden algebra $sl(4, \bf R)$, while latter one has the hidden algebra $h^{(3)}$, thus, becoming Lie-algebraic operators. Both operators can be extended to (quasi)-exactly-solvable operators which have an extra integral and admit a separation of one variable. Interestingly, both (quasi)-exactly-solvable operators lead to the {\it same} (quasi)-exactly-solvable potentials in the space of relative distances.

The above formalism admits a natural generalization to the case of arbitrary $d$ dimensional three bodies. The Laplace-Beltrami operator remains unchanged, the effective potential (\ref{HLB3}) is changed but not dramatically; (quasi)-exactly-solvable integrable models continue to exist. It will be presented elsewhere.

\section*{Acknowledgments}

A.V.T. is thankful to University of Minnesota, USA for kind hospitality extended to him where this work was initiated and IHES, France where it was completed. He is deeply grateful to I.E.~Dzyaloshinsky (Irvine), T.~Damour, M.~Kontsevich (Bures-sur-Yvette) and Vl.G.~Tyuterev (Reims) for useful discussions and important remarks.
A.V.T. is supported in part by the PAPIIT grant {\bf IN108815} and CONACyT grant {\bf 166189}~(Mexico).
W.M. was partially supported by a grant from the Simons Foundation (\# 208754 to Willard Miller, Jr.).
M.A.E. is grateful to ICN UNAM, Mexico for the kind hospitality during his visit, where a part of the research was done, he was supported in part by DGAPA grant {\bf IN108815} (Mexico) and, in general, by  CONACyT grant {\bf 250881}~(Mexico) for postdoctoral research.

\section*{Appendix: non-equal masses}

Consider the general case of the particles located at points ${\bf r}_1,{\bf r}_2,{\bf r}_3$
of masses $m_1,m_2,m_3$, respectively.
Then the operator (\ref{addition3-3rho}) becomes (in terms of the relative coordinates $\rho_{ij}=r_{ij}^2$):
\[
  {\De_R}'(\rho_{ij})\ =\
  \frac{2}{\mu_{13}} \rho_{13}\, \pa_{\rho_{13}}^2 +
  \frac{2}{\mu_{23}} \rho_{23}\, \pa_{\rho_{23}}^2 +
  \frac{2}{\mu_{12}} \rho_{12}\,\pa_{\rho_{12}}^2 +
\]
\[
  \frac{2(\rho_{13} + \rho_{12} - \rho_{23})}{m_1}\pa_{\rho_{13}\rho_{12}} +
  \frac{2(\rho_{13} + \rho_{23} - \rho_{12})}{m_3}\pa_{\rho_{13}\rho_{23}} +
  \frac{2(\rho_{23} + \rho_{12} - \rho_{13})}{m_2}\pa_{\rho_{23}\rho_{12}} +
\]
\begin{equation}
\label{addition3-3r-M}
  \frac{3}{\mu_{13}} \pa_{\rho_{13}} +
  \frac{3}{\mu_{23}} \pa_{\rho_{23}} +
  \frac{3}{\mu_{12}} \pa_{\rho_{12}}\ ,
\end{equation}
where
\[
   \frac{1}{\mu_{ij}}\ =\ \frac{m_i+m_j}{m_i m_j}\ ,
\]
is reduced mass for particles $i$ and $j$; it is in agreement with  (\ref{addition3-3rho})
for $m_1=m_2=m_3=1$ or, equivalently, $\mu_{ij}=1/2$. This operator has the same algebraic structure
as ${\De_R}(\rho_{ij})$ but lives on a different manifold in general. It can be rewritten in terms
of the generators of the maximal affine subalgebra $b_4$ of the algebra $sl(4,{\bf R})$,
see (\ref{sl4R}), c.f. (\ref{HRex}). The determinant of the contravariant metric tensor is
\[
  D_m\ =\ \det g^{\mu \nu}\ =\ 2\,\frac{m_1+m_2+m_3}{m_1^2m_2^2m_3^2} \times
\]
\begin{equation}
\label{gmn33-rho-det-M}
 \left(m_1m_2\rho_{12}+m_1m_3\rho_{13}+m_2m_3\rho_{23}\right)
                     \left(2\rho_{12}\rho_{13} + 2 \rho_{12}\rho_{23} + 2 \rho_{13}\rho_{23}-\rho_{12}^2- \rho_{13}^2 - \rho_{23}^2\right) \ ,
\end{equation}
and is positive definite. It is worth noting a remarkable factorization property
\[
D_m\ =\ 2\frac{m_1+m_2+m_3}{m_1^2m_2^2m_3^2} \,(m_1 m_2 r_{12}^2+m_1 m_3 r_{13}^2+m_2 m_3 r_{23}^2)\ \times
\]
\[
(r_{12}+r_{13}-r_{23})(r_{12}+r_{23}-r_{13})(r_{13}+r_{23}-r_{12})(r_{12}+r_{13}+r_{23})\ =
\]
\[
   =\ 32\, \frac{m_1+m_2+m_3}{m_1^2m_2^2m_3^2}\, P' \ S^2_{\triangle}\ ,
\]
where $P'=m_1 m_2 r_{12}^2+m_1 m_3 r_{13}^2+m_2 m_3 r_{23}^2$ - the weighted sum of squared of sides of the interaction triangle and $S_{\triangle}$ is their area. Hence, $D_m$ is still proportional to  $S_{\triangle}^2$, c.f. (\ref{gmn33-rho-det-gamma}).

Making the gauge transformation of (\ref{addition3-3r-M}) with determinant (\ref{gmn33-rho-det-M}) as the factor,
\[
         \Gamma \ = \ D_m^{-1/4}\ ,
\]
we find that
\begin{equation}
         D_m^{1/4}\, {\De_R}'(\rho_{ij})\,D_m^{-1/4} \ =
        \  \De'_{LB}(\rho_{ij}) - \tilde V_m \ ,
\label{HLB3M}
\end{equation}
is the Laplace-Beltrami operator with the effective potential
\[
{\tilde V_m} \ =\ \frac{3(m_1+m_2+m_3)}{8 \left(m_1 m_2 r_{12}^2+m_1 m_3 r_{13}^2+m_2 m_3 r_{23}^2\right)}\ + \ \frac{\left(m_1 m_2 r_{12}^2+m_1 m_3 r_{13}^2+m_2 m_3 r_{23}^2\right)}
  {2 m_1 m_2 m_3\left(\rho _{12}^2+\rho _{13}^2+\rho _{23}^2 -2 \rho _{12} \rho _{13}-
                     2 \rho _{12} \rho _{23}-2 \rho _{13} \rho _{23}\right)}\ .
\]
The Laplace-Beltrami operator plays a role of the kinetic energy of three-dimensional quantum particle moving in curved space. It seems evident the existence of (quasi)-exactly-solvable problems with such a kinetic energy, see e.g. \cite{Crandall:1985} as for the example of exactly-solvable problem.

\end{document}